\documentstyle[psfig,12pt,aaspp4]{article}

\lefthead{Barger et al.}

\begin{document}
\title{Constraints on the Early Formation of Field Elliptical Galaxies}
\author{A.~J.~Barger,\altaffilmark{1,}\altaffilmark{2}
L.~L.~Cowie,\altaffilmark{1}
N.~Trentham,\altaffilmark{3}
E.~Fulton,\altaffilmark{1,}\altaffilmark{2}
E.~M.~Hu,\altaffilmark{1}
A.~Songaila,\altaffilmark{1}
D.~Hall\altaffilmark{1}
}

\altaffiltext{1}{Institute for Astronomy, University of Hawaii, 
2680 Woodlawn Drive, Honolulu, Hawaii 96822, USA}
\altaffiltext{2}{Visiting Astronomer, Canada-France-Hawaii Telescope,
operated by the National Research Council of Canada, the Centre
National de la Recherche Scientifique of France, and the University
of Hawaii}
\altaffiltext{3}{Institute of Astronomy, University of Cambridge, 
Madingley Road, Cambridge CB3 0HA, UK}

\begin{abstract}

We present the results of an $HK^\prime$ wide-field survey 
encompassing the Hubble Deep Field and its Flanking Fields. 
Our wide-field survey provides uniform coverage of an area 
$61.8$\ arcmin$^2$ to a depth equivalent to $K=20.1$ at
$5\,\sigma$. We have also imaged
the Hubble Deep Field in $HK^\prime$, providing uniform coverage of an
area $7.8$\ arcmin$^2$ to a depth equivalent to $K=21.2$ at $5\,\sigma$. 
Using these data in combination with new deep UH 8K $V$ and $I$ imaging
obtained on the CFHT, we find only a small population of
objects with colors redder than an 
equivalent $I-K=4$, the color expected for an evolved elliptical at $z>1$. 
We infer that only a fraction of the local field elliptical
population with $M_K<-23.4$ could have formed in single bursts at high 
redshift.

\end{abstract}


\section{Introduction}
\label{intro}

The study of galactic evolution is a key area of observational cosmology
since it is central to understanding the nature of galaxies and 
to probing the cosmological geometry. With observational data now revealing
normal galaxies at redshifts as high as $z\sim 5$, we should be close to
being able to locate or rule out the existence of a substantial
high redshift field elliptical galaxy population.
These old systems would have formed at the earliest times, and hence their
existence would place very stringent constraints on galaxy and structure
formation.

How elliptical galaxies form remains a topic of debate. The standard view 
(e.g.\ \markcite{eggen62}Eggen, Lynden-Bell \& Sandage 1962;
\markcite{tinsley76}Tinsley \& Gunn 1976)
has been that elliptical galaxies
formed the bulk of their stars in a single burst of star formation at high 
redshift, after which their stellar populations evolved passively.
This monolithic collapse model for elliptical formation is in marked contrast
with the hierarchical merging scenario for galaxy formation and 
evolution in which massive galaxies form at all redshifts from
the gradual merging of smaller galaxies (e.g.\ 
\markcite{white91}White \& Frenk 1991;
\markcite{kauffmann93}Kauffmann, White \& Guiderdoni 1993; 
\markcite{baugh96}Baugh et al.\ 1996).
The monolithic collapse model appears to be supported by extensive studies
of elliptical galaxies in rich cluster environments to $z>1$. 
Observational studies of the color evolution in cluster early-type galaxies
have demonstrated that the changes with redshift are consistent with a 
model in which stellar populations formed at high redshift, $z_f>2$, and
subsequently evolved passively
(e.g.\ \markcite{alfonso93}Arag{\'o}n-Salamanca et al.\ 1993).
A similar $z_f>2$ constraint is obtained from
the observed small scatter in the color-magnitude relations in distant 
clusters (e.g.\ \markcite{ellis97}Ellis et al.\ 1997;
\markcite{stan98}Stanford, Eisenhardt \& Dickinson 1995, 1998), and from
the evolution of the Mg\,b versus $\sigma$ relation
(e.g.\ \markcite{bz97}Ziegler \& Bender 1997) and the fundamental plane relation
(e.g.\ \markcite{vDF96}van Dokkum \& Franx 1996;
\markcite{kelson97}Kelson et al.\ 1997). However,
\markcite{kauffcharlot98}Kauffmann \& Charlot (1998) have argued that in the 
rich cluster environment hierarchical elliptical galaxy evolution may appear 
indistinguishable from that predicted by the passive evolution scenario. 
In their model, rich clusters at high redshift correspond to regions where the 
very earliest star formation occurred, and hence the star formation and merging 
histories of galaxies in these clusters have progressed much further than 
have those of field galaxies. Thus, in 
observing apparently similar clusters of galaxies, observers may be 
selecting only the oldest galaxies at any particular redshift.

Because the passive evolution scenario for elliptical galaxies predicts 
a population of $z>1$ galaxies with ultrared optical-near-IR colors,
this scenario can be sensitively tested in the field environment 
with very deep optical and near-infrared imaging surveys.
Low-redshift galaxies with significant star formation can mimic
the very red colors of an evolved elliptical if they are greatly reddened 
by dust, but
the number of ultrared objects observed gives an {\em upper} limit on the
comoving number density of passively evolving elliptical galaxies, provided
that there does not exist a population of completely dust-obscured
high redshift objects. 
Imaging surveys have only recently become deep enough in both the optical
and the near-IR bands to test for the presence of ultrared galaxies at $z>1$.

\markcite{cowie94}Cowie et al.\ (1994) presented a very deep $K$-band 
survey of four small fields of total area $5.9$\ arcmin$^2$ 
which they later followed up spectroscopically
(\markcite{songaila94}Songaila et al.\ 1994). From their data
they found that the space density of extremely red objects
having $I-K>4$, the expected color for old stellar populations at $z>1$, 
implied that at most ten per cent of the total local 
galaxy population could have formed in early single-starburst events.
A comparison of their $K<20.9$ sample with a wider, brighter
sample having $K<17.8$ also showed a scarcity of very red objects
which could lie at redshifts $z>1$. Thus, Cowie et al.\ suggested 
that either galaxies are less luminous at these higher redshifts or
the mix of galaxy types is much bluer than expected. Recently,
\markcite{hogg97}Hogg et al.\ (1997) and 
\markcite{moustakas97}Moustakas et al.\ (1997) analyzed a few very 
deep, small-area fields in $K$. Hogg et al.\
imaged two subfields ($\sim 1$\ arcmin$^2$ total area) in the 
Hubble Deep Field (hereafter HDF; \markcite{williams96}Williams et al.\ 1996)
with the Near Infrared Camera on Keck to a 90 per cent completeness
limit of $K=23$. Most of their sources had colors consistent with either
old stellar populations observed at $0<z<1$ or younger populations observed 
over a wide range of redshifts. However, they also found four sources which 
had very red colors, $I_{814}-K>4$.
\markcite{moustakas97}Moustakas et al.\ presented colors
for very faint galaxies ($K<23$) in two fields covering an area 
$\sim 2$\ arcmin$^2$. They found a small population of
galaxies in their sample whose infrared bright ($I-K>4$) 
but relatively blue optical ($V-I\lesssim 2.5$) colors could not
be reproduced with any of their standard models. Moustakas et al.\
suggested three possible scenarios to explain these `red outliers',
namely (i)\,a universal event of secondary bursts in $z\sim 3$ passively
evolving elliptical galaxies, (ii)\,a combination of very
low metallicity and substantial reddening in $z<1.5$ galaxies,
or (iii)\,the existence of dusty,
old populations within the galactic nuclei of normal galaxies.
In their preliminary follow-up analysis,
\markcite{moustakas98}Moustakas et al.\ (1998) presented {\it HST} WFPC2
and NICMOS morphologies for some of their red outliers. These images
showed the objects to be compact and asymmetric and hence unlikely to be 
relaxed systems. Thus, these authors now prefer the explanation that the 
red outlier population consists of very dusty $z>1.2$ systems containing 
bright nuclei powered by either AGNs or starbursts.
In the present paper we find that such objects are much
rarer than was measured by Moustakas et al.

The results of the above three surveys, as well as those from recent 
$K$-band HDF imaging by Dickinson et al.\ (in preparation), 
were subsequently analyzed by
\markcite{zepf97}Zepf (1997) in an effort to determine the dominant formation
mechanism for elliptical galaxies.
Zepf compared the observed surface densities of the ultrared galaxies to a
given magnitude limit with their expected surface densities, assuming a 
passive formation model in which all early-type galaxies form in short bursts 
at high redshift.
He concluded that there is a strong deficit of galaxies with extremely red 
colors in the observed fields (therefore ruling out models where typical
ellipticals are fully assembled and have formed all of their stars by $z>5$),
and thus that most ellipticals must have formed at $z<5$ through merging and 
associated starbursts. 

Using the HDF dataset of Dickinson et al., 
\markcite{frances98}Franceschini et al.\ (1998) analyzed the
spectral energy distributions of morphological ellipticals in the HDF
to date the dominant stellar populations. They found that the majority
of the bright early-types were at $z\lesssim 1.3$ and had colors which indicated
ages of typically $1.5-3$\ Gyr. The sudden disappearance of 
objects at $z>1.3$, when the objects should have been visible
in a luminous star-formation phase, is explained by 
\markcite{frances98}Franceschini et al.\ 
as being due to the fact that during the first few Gyrs of the galaxy's 
lifetime the objects are undergoing dust-enshrouded merger-driven starbursts.

Provided that high redshift ellipticals are not completely obscured by dust,
deep infrared surveys can detect their presence and place constraints on
their surface density. Existing
deep $K$-band surveys have concentrated on small areas due to
the small format of the IR arrays. However, with the development of large-format
IR arrays such as the HAWAII ($1024\times 1024$) HgCdTe devices, it is now
feasible to conduct deep near-infrared surveys over wider fields.
In this paper we present a wide-field $HK^\prime$ survey of 
the HDF and its Flanking Fields (FFs) with corresponding $V$ and $I$ colors.
The primary goal of our imaging survey is to compile a well-defined sample of
near-infrared selected galaxies over a large area to place constraints on 
the population of ultrared galaxies. As a complement to this wide-field 
survey, we have also obtained very deep $HK^\prime$-band observations of the HDF 
itself.

In \S\ref{data} we describe our new data sample and the construction of
our photometric catalogs.
In \S\ref{ultrared} we use our imaging data to study galaxy colors
and the effects of dust, and
in \S\ref{disc} we use our data to put constraints on the high redshift 
field elliptical population.
\S\ref{summary} summarizes our results. 

\section{Data}
\label{data}

\subsection{Observations}
\label{obs}

To complement the multi-color data of the HDF, we have obtained
deep follow-up $HK^\prime$ observations of the HDF and its FFs
using the University of Hawaii 2.2-m telescope and
the Canada-France-Hawaii 3.6-m telescope (CFHT). We have also obtained
deep $V$ and $I$-band observations with the CFHT over a much larger area
and Keck LRIS $B$-band observations in a strip across the center of the 
$HK^\prime$ wide-field image.

The wide-field near-infrared images were obtained with the UH 2.2-m on the 
nights of UT 1997 April 17-22 with the University of Hawaii QUick Infrared 
Camera (QUIRC; \markcite{hodapp96}Hodapp et al.\ 1996). QUIRC utilizes 
a $1024\times 1024$ pixel HgCdTe Astronomical Wide Area Infrared
Imaging (HAWAII) array produced by the Rockwell
International Science Center. At the f/10 focal ratio, the
field of view of QUIRC is $193\times 193$~arcsec square 
with a scale of $0.1886$~arcsec~pixel$^{-1}$. 
The observations were made using a notched $HK^\prime$ filter with a 
central wavelength of $1.8\micron$ (Wainscoat \& Cowie, in preparation).
The filter covers the longer wavelength region of the $H$-band and the
shorter wavelength region of the $K$-band (roughly the $K'$ filter).
Because of its broad bandpass and low sky background, this filter is
extremely fast and is roughly twice as sensitive as the $H$, $K'$, or $K_s$
filters. We covered the full area imaged with the {\it HST} using a
$3\times 3$ mosaic having a $20$~arcsec overlap. Thirteen 
spatially shifted short exposures (around 100 sec each
depending on the background level) of step-size
$5-20$~arcsec in all directions were obtained at
each of the nine pointings within the full mosaic, and the mosaic pattern
was repeated several times during the run. This observing
strategy enabled us to use the data images for the subsequent sky 
subtraction and flat-fielding procedures. Each object is
sampled by a different array pixel on each exposure, and
the signal-to-noise ratio should improve as the square root of the number
of frames. The dark current is insignificant in the QUIRC data. 
The data were processed using median sky flats generated from the
dithered images and calibrated from observations of UKIRT faint
standards taken on the first three nights when the sky was photometric.
The FWHM on the total composite image was $0.8$~arcsec.
The magnitudes were measured
in $3.0$~arcsec diameter apertures and corrected
to total magnitudes following the procedures of
\markcite{cowie94}Cowie et al.\ (1994).

\begin{figure}
\centering
\hspace*{0in}\psfig{figure=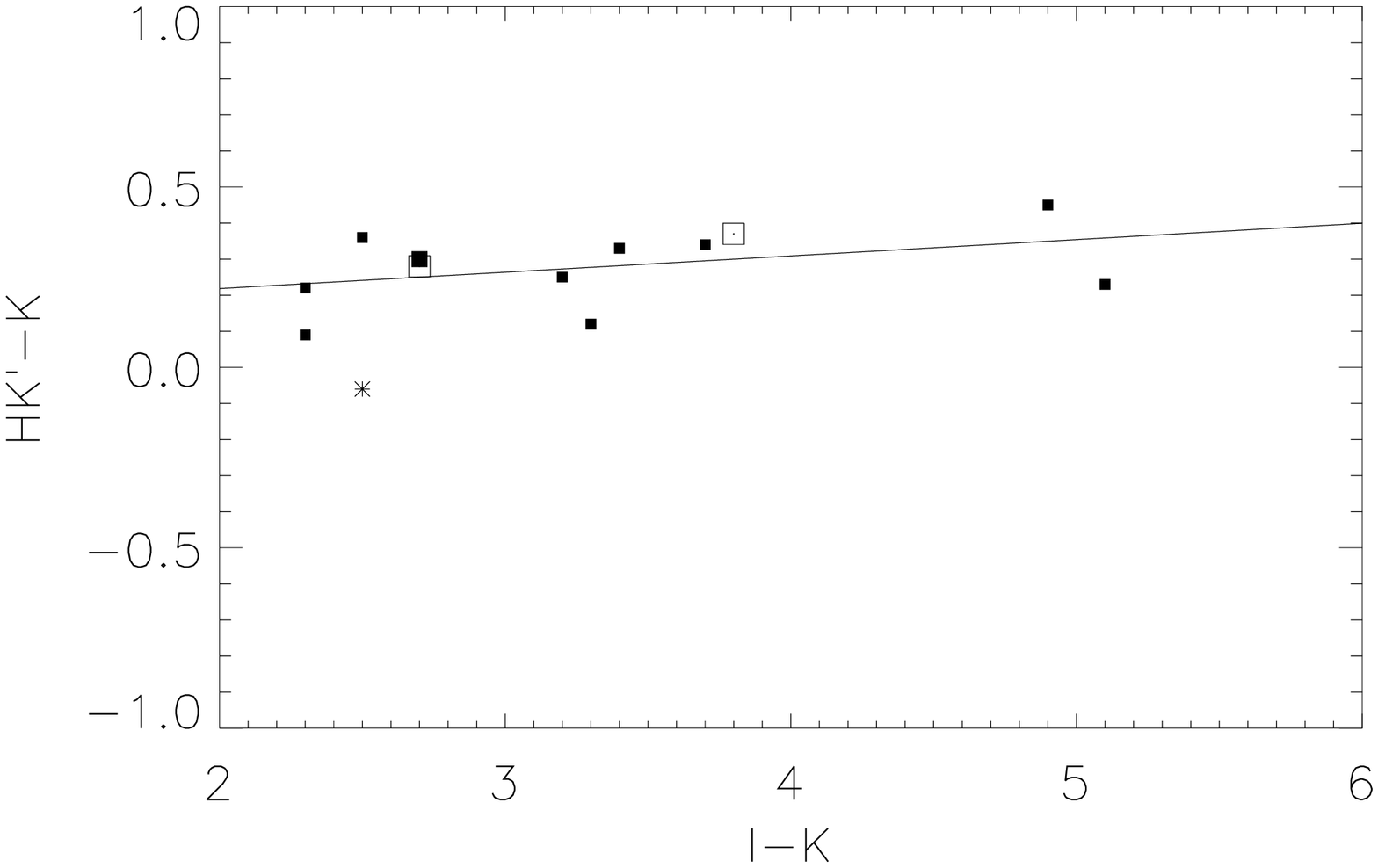,width=4.0in,height=3.0in,angle=90}
\caption{$HK^\prime - K$ versus $I - K$ for objects
in the Hawaii Deep Survey Field SSA13: stars (star symbol; only one) and
galaxies with magnitudes
$K<18$ (open squares), $K=18-19$ (large closed square; only one),
and $K=19-20$ (small closed squares).
The galaxy color equation obtained from these data is
$HK^\prime - K = 0.13 + 0.05 (I-K)$.
\label{hkkik}}
\end{figure}

Since we will want to compare our $HK^\prime$ measurements with 
previous $K$-band data, we need a conversion between the two bands. 
From $K$, $HK^\prime$, and $I$ images of galaxies in the SSA13
field, we find the empirical relation 
$HK^\prime - K = 0.13 + 0.05 (I-K)$. Figure~\ref{hkkik} shows
$HK^\prime - K$ versus $I-K$ colors for a range of redshifts and 
galaxy types in the SSA13 field. The offset between the galaxies 
and the lone star in the figure reflects the different shapes 
of stars and galaxies in the wavelength region
$1.5-2.4$\micron. Stars rise to shorter wavelengths whereas galaxies
are flatter, so even blue galaxies are offset from stars.
Since the coefficient of the
$I-K$ term is small, it suffices for us to use a median value for 
$I-K$ in this relation, and we therefore obtain $HK^\prime - K = 0.30$.


Deep imaging of the HDF itself was obtained in the $HK^\prime$ filter with 
QUIRC on the UH 2.2-m on the nights of UT 1996 February 6-8 and
with QUIRC on the CFHT on the nights of UT 1996 5-8 April. 
The CFHT image was registered to the UH 2.2-m data and a noise-weighted
addition was performed. The net image quality is $\sim 1''$ FWHM.
The noise in each field was estimated by measuring corrected 
$3$~arcsec diameter magnitudes at a number of blank sky positions.
The $2\,\sigma$ limit for the wide-field image is $HK^\prime=21.4$ and 
for the deep image is $HK^\prime=22.5$.

The $V$ and $I$ optical images were obtained on the nights of 
UT 1997 April 3-8 with the CFHT
and the University of Hawaii 8K CCD Mosaic Camera, a $2\times 4$ 
array of $4096\times 2048$ CCDs built by Metzger, Luppino, and 
Miyazaki. The image scale is $0.21$~arcsec~pixel$^{-1}$.
The observations were made through Kron-Cousins $I$ and Johnson $V$
filters and comprise twenty-two 1200 sec $I$-band exposures (8.3\,hrs)
and thirty-three 1200 sec $V$-band exposures (11\,hrs) under conditions
of mixed transparency. Net image quality is $\sim 0.9''$ FWHM in $I$ and
$\sim 1.0''$ FWHM in $V$. The UH 8K data were processed 
in the standard way using the imcat 
data reduction routines written by Nick Kaiser (\markcite{kaiser97}1997). 
The data were calibrated to total Kron-Cousins $I$ and Johnson $V$
magnitudes using shallower images obtained under
photometric conditions with the UH 2.2-m telescope. The $2\,\sigma$ limit
for the $I$-band image is $I=24.8$ and for the $V$-band image is $V=25.8$.

\subsection{Photometric Measurements and Catalogs}
\label{catalogs}

Objects were detected and measured on all the images
using the SExtractor package written 
by \markcite{bertin96}Bertin \& Arnouts (1996).
Before extraction the data were convolved with a gaussian chosen to
have a FWHM approximately equal to the seeing.
Objects were extracted if they contained at least
25 connected pixels each with a signal above a threshold of $1\,\sigma$.
The photometric measurements were then made on the unconvolved images
using $3$~arcsec diameter apertures, with the photometric
calibration described in \S\ref{obs}, and corrected
to $6$~arcsec diameter magnitudes (\markcite{cowie94}Cowie et al.\ 1994). 
All magnitudes presented in this paper are given as these near
total magnitudes.

We determined the incompleteness of the wide-field $HK^\prime$ catalog by 
measuring objects in the wide-field image at the positions of the deep
near-infrared catalog. By comparing the fraction of
galaxies already included in the wide-field $HK^\prime$ 
sample with those measured using the deep $HK^\prime$ catalog as a 
function of near-infrared magnitude, we
found that the wide-field near-infrared catalog became more than
50 per cent incomplete at $HK^\prime\gtrsim 21.3$.
We further checked the incompleteness of the wide-field near-infrared 
catalog, and determined the incompleteness of the deep $HK^\prime$ field,
by constructing $I$ and $V$ samples with SExtractor (as above). We
measured photometry at the optical positions on the near-infrared images.
We compared the fraction of galaxies already included in the $HK^\prime$ sample 
with those measured using the optically-selected catalogs as a function of 
near-infrared magnitude and found a consistent incompleteness level for
the wide-field near-infrared catalog with what we found above. 
The deep catalog became more than 50 per cent incomplete at 
$HK^\prime\gtrsim 22.3$.

To determine the J2000 coordinates for our $HK^\prime$-selected sample, 
we first obtained the plate scale and the orientation angle
for the QUIRC camera using our wide-field $HK^\prime$ image. 
We chose to use the APM eo674 plate, where our field is slightly
more centered. From a fit to 33 APM objects ($17<V<22.5$) covering our field,
we found a plate scale of 0.189~arcsec~pixel$^{-1}$ and an orientation angle 
of 0.636\ deg North through East, both of which are consistent with previous 
determinations for the QUIRC camera (the orientation angle was previously
determined to be 0.728\ deg for B1950). The dispersion of the fit was 
0.32~arcsec 
in both the RA and Dec directions, giving a total error of 0.45~arcsec.
This error is comparable to the expected astrometric errors
in the APM catalog. As a check, we compared the RA and Dec
positions in the two APM plates covering the region (eo1427 and
eo674) and found an rms error between the two plates of 0.73~arcsec. 
The average offsets of 1.8~arcsec in RA and 1.5~arcsec in Dec between 
the positions measured from the
two plates are consistent with the APM errors which fully dominate the 
error budget. 
(Our field lies close to the edge of the eo1427 plate where the APM solution 
may be worse.) Futhermore, the offsets are consistent with the
advertised accuracies of the APM absolute astrometry. No systematic
effects were visible.

The offsets to the VLA system, determined using the 5 radio sources
in \markcite{fomalont97}Fomalont et al.\ (1997) which have
the smallest RA and Dec errors (0.2~arcsec) from their identified 
$I<22$ optical counterparts, were found to be 1.68~arcsec in the 
RA direction and -0.59~arcsec in the Dec direction,
consistent with the expected APM absolute errors. 
The RMS errors of 0.16~arcsec in the RA direction and 
0.11~arcsec in the Dec direction are consistent with the radio 
source errors.

The wide-field catalog of $HK^\prime$-selected objects whose
centers fall within an area $\sim7.9\times7.9$\ arcmin$^2$,
the region over which the coverage is uniform, has been posted to a
web page linked to the Hubble Deep Field Clearing House page at STScI.
For each object we give an RA and Dec catalog number, a
J2000 RA and Dec tied to the VLA coordinates of
\markcite{fomalont97}Fomalont et al.\ (1997) (described above), and our
$HK'$, Kron-Cousins $I$, Johnson $V$, and,
where available, $B$ total magnitudes.
Ground-based spectroscopic survey observations of galaxies in the HDF have been
taken with the Keck telescope and either published or made publicly
available by several groups
(\markcite{cohen96}Cohen et al.\ 1996;
\markcite{steidel96}Steidel et al.\ 1996;
\markcite{lowenthal97}Lowenthal et al.\ 1997;
\markcite{phillips97}Phillips et al.\ 1997;
\markcite{songaila97}Songaila 1997).
These spectroscopic redshifts are included in the catalog.

\subsection{Star-Galaxy Separation}
\label{starsep}

Stars are cleanly separated from galaxies in the $B-I$ versus 
$I-K$ plane. The $I-K$ colors are primarily a 
redshift diagnostic. As the redshift increases the ultraviolet part
of the galaxy spectrum shifts into the $I$-band, and the $I-K$ 
colors become progressively redder. On the other hand, $B-I$ colors 
are primarily a diagnostic of galaxy morphology. The empirical 
star-galaxy separation line
has been found to be $(B-I) - 2.5 (I-K) = -2.0$ 
(\markcite{kron80}Kron 1980; \markcite{gardner92}Gardner 1992; 
\markcite{huang97}Huang et al.\ 1997).
For the objects in our data sample for which we have 
$B$-band data ($\sim 50$ per cent), this line, converted to $HK^\prime$, 
also provides an appropriate star-galaxy separation.
We determine a corresponding separation line in 
$V-I$ versus $I-HK^\prime$ that reproduces the above star or galaxy 
classifications $[(V-I) - 2.0 (I-HK^\prime) = -1.1]$ and apply it to 
classify the remaining objects in our wide-field sample 
(see Fig.~\ref{necolcol}).

\section{Galaxy Colors in the Near-Infrared Samples}
\label{ultrared}

\begin{figure}
\centering
\hspace*{0in}\psfig{figure=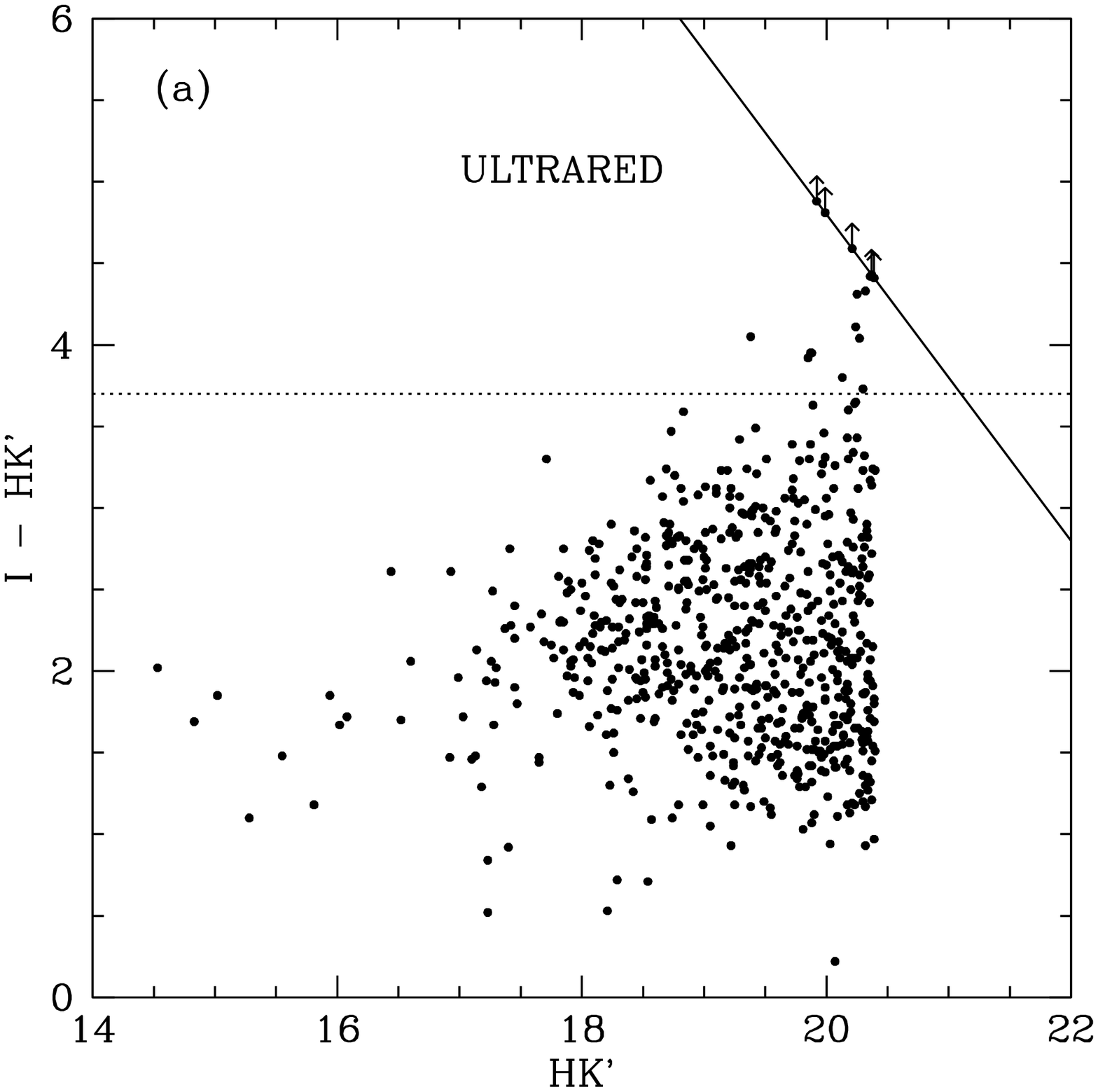,width=3.0in,height=3.0in}
\hspace*{0in}\psfig{figure=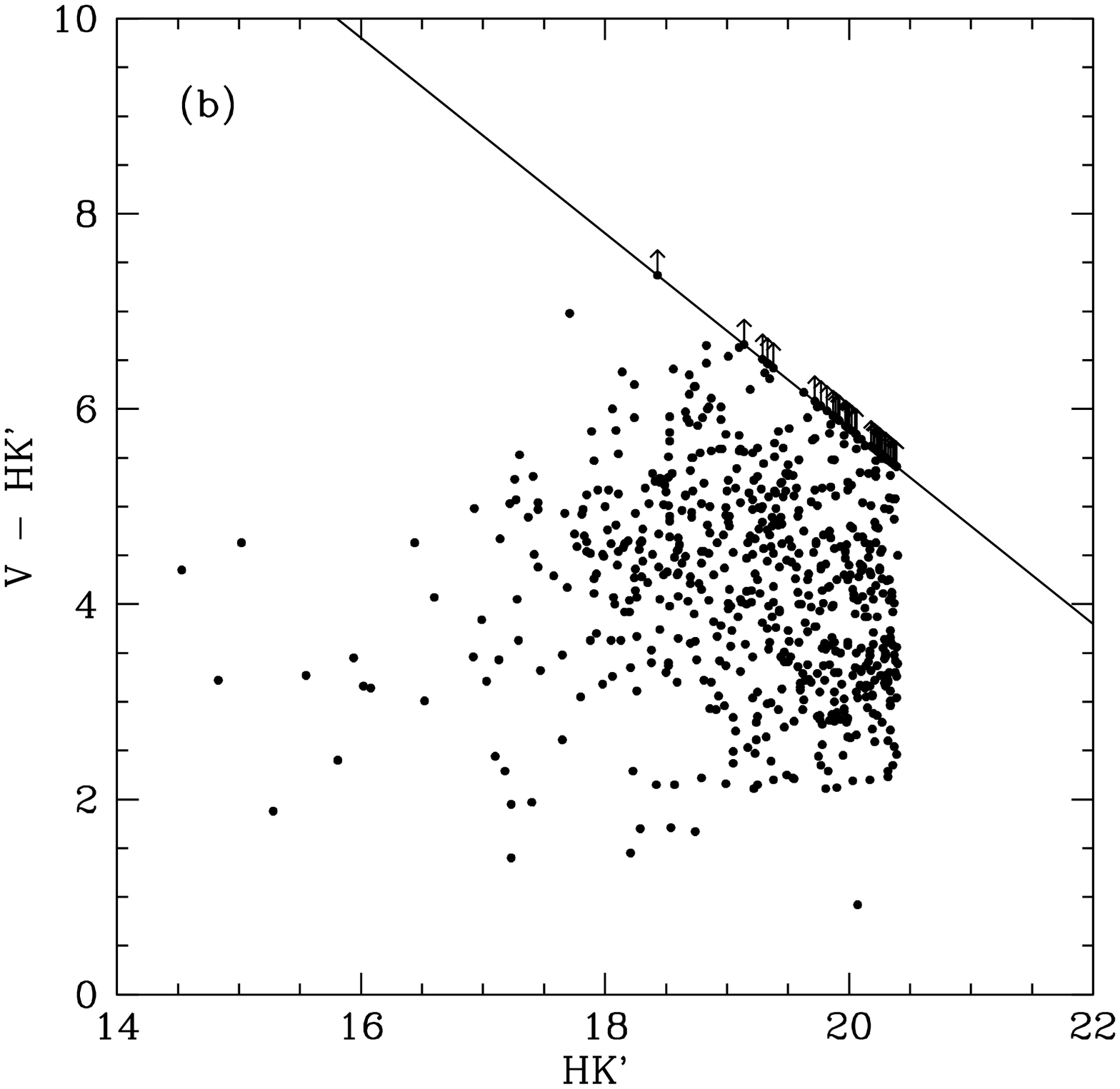,width=3.0in,height=3.0in}
\hspace*{0in}\psfig{figure=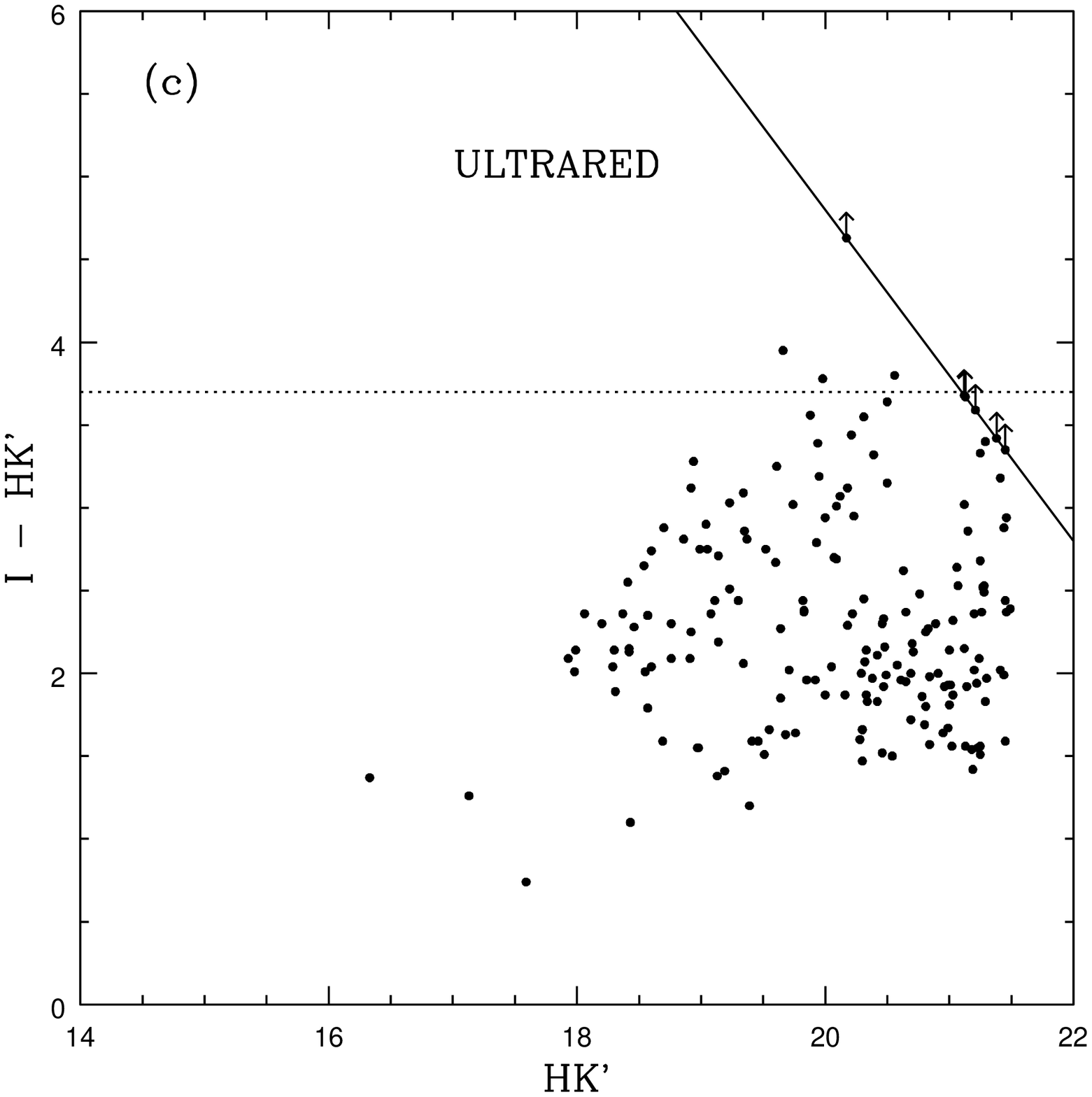,width=3.0in,height=3.0in}
\hspace*{0in}\psfig{figure=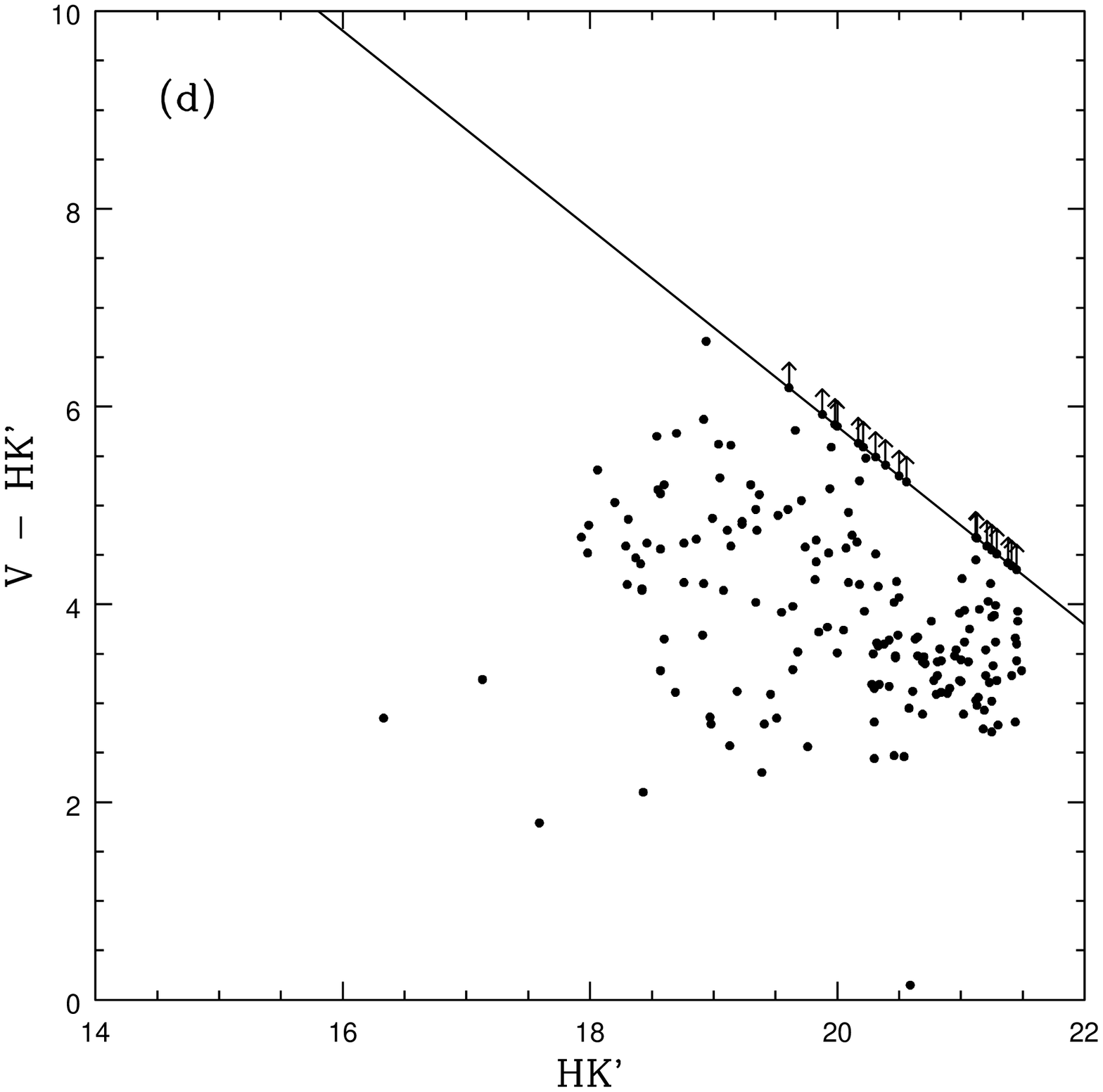,width=3.0in,height=3.0in}
\caption{Color-magnitude diagrams for the $5\,\sigma$
$HK^\prime$-selected sample of objects (stars and galaxies):
(a) $I-HK^\prime$ versus $HK^\prime$ and
(b) $V-HK^\prime$ versus $HK^\prime$ for the wide-field sample;
(c) $I-HK^\prime$ versus $HK^\prime$ and
(d) $V-HK^\prime$ versus $HK^\prime$ for the deep sample.
The solid lines are $2\,\sigma$ upper limits on the colors.
The dotted lines in (a) and (c)
mark the lower limit for ultrared galaxies ($I-HK^\prime > 3.7$).
The upward pointing arrows represent objects which were undetected at
the $2\,\sigma$ level in $I$ (a, c) or $V$ (b, d).
\label{colmag}}
\end{figure}

In Fig.~\ref{colmag} we show the $I-HK^\prime$ versus $HK^\prime$ 
and $V-HK^\prime$ versus $HK^\prime$
color-magnitude diagrams (both stars and galaxies) for the
$5\,\sigma$ near-IR selected samples in
our wide-field and deep HDF images. Only
five of the 664 objects in the $5\,\sigma$ wide-field sample were not 
detected at or above the $2\,\sigma$ level in $I$, and
41 were not detected at or above the $2\,\sigma$ level in $V$. 
For the deep HDF image, six of the 184 objects in the $5\,\sigma$ 
near-IR selected sample were not detected at or above the $2\,\sigma$ level 
in $I$, and 18 were not detected at or above the $2\,\sigma$ level in $V$.
We assign these objects
$2\,\sigma$ limiting magnitudes for the following analysis, and we mark
their $V-HK^\prime$ and $I-HK^\prime$ colors as lower limits on the plots 
with an upward pointing arrow. The solid lines show the $2\,\sigma$ 
magnitude limits on the $I-HK^\prime$ or $V-HK^\prime$ colors,
and the dotted lines in the two $I-HK^\prime$ versus $HK^\prime$ 
figures indicate our adopted definition of $I-HK'>3.7$ for the ultrared 
galaxy population, which we shall discuss in detail below.
 

\begin{figure}
\centering
\hspace*{0in}\psfig{figure=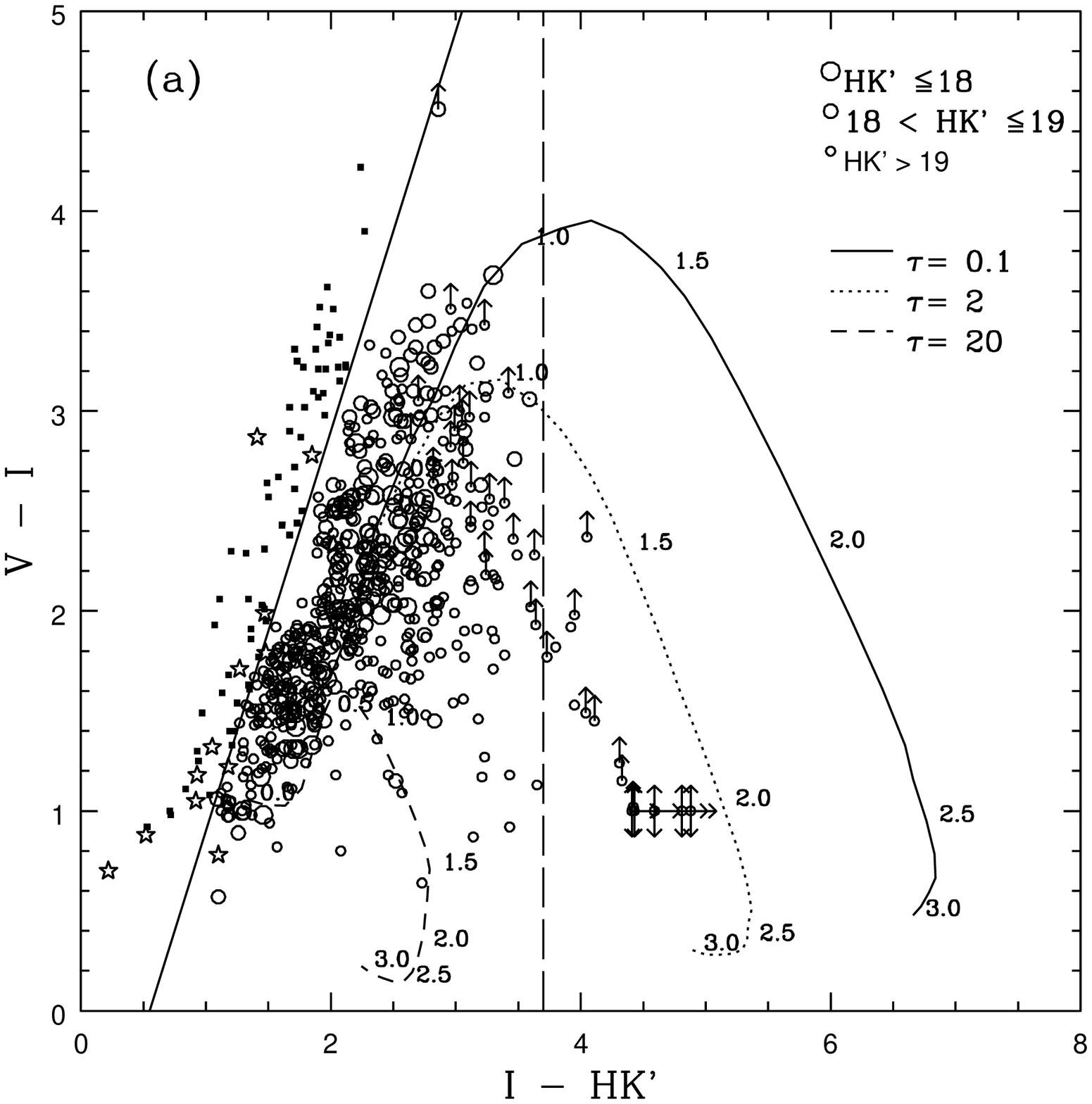,width=3.0in,height=3.0in}
\hspace*{0in}\psfig{figure=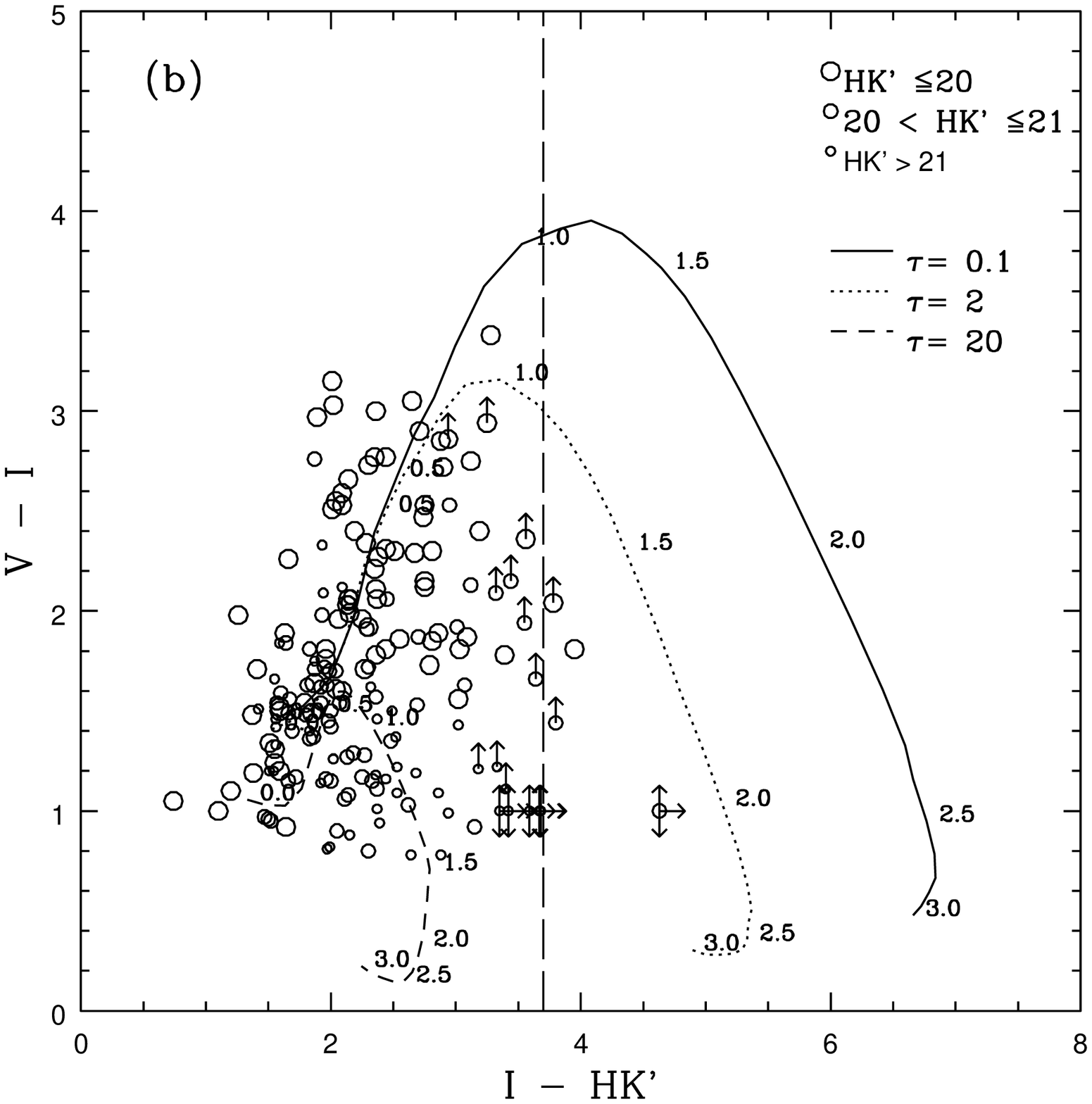,width=3.0in,height=3.0in}
\caption{$V-I$ versus $I-HK^\prime$ color-color diagrams for
(a) the wide-field $HK^\prime\le 20.4$ sample and
for (b) the deep HDF $HK^\prime\le 21.5$ sample.
Bruzual \& Charlot (1996) pure K-correction model tracks folded
with the appropriate $V$, $I$, and $HK^\prime$ filters are illustrated
from $0<z<3$, assuming a common epoch of formation $z_f=5$.
The model tracks have been labeled with a few representative redshifts
at which the various colors would be observed. In (a) the
adopted star-galaxy separation line determined in \S 2.3
is indicated by a solid line.
The star symbols are definite stars identified
either by spectroscopy or from the HST images. The solid squares
are objects assumed to be stars based on their location above
the star-galaxy separation line in the color-color plane.
All objects classified as stars were removed from the wide-field sample
for the analysis. No stars were removed from the deep sample.
The vertical dashed lines in the figures
illustrate the ultrared dividing line, $I-HK^\prime>3.7$.
\label{necolcol}}
\end{figure}

\begin{figure}
\centering
\hspace*{0in}\psfig{figure=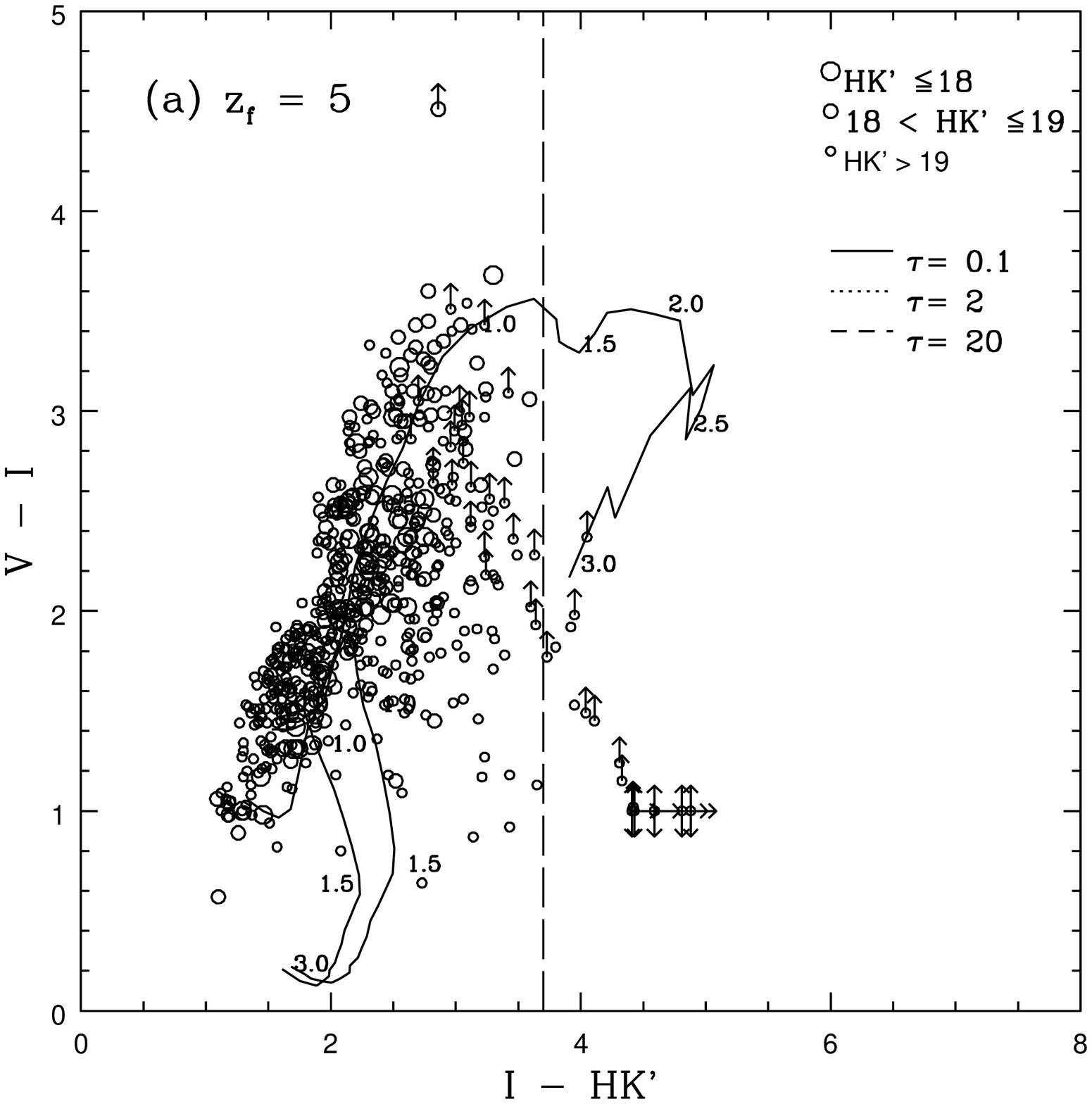,width=3.0in,height=3.0in}
\hspace*{0in}\psfig{figure=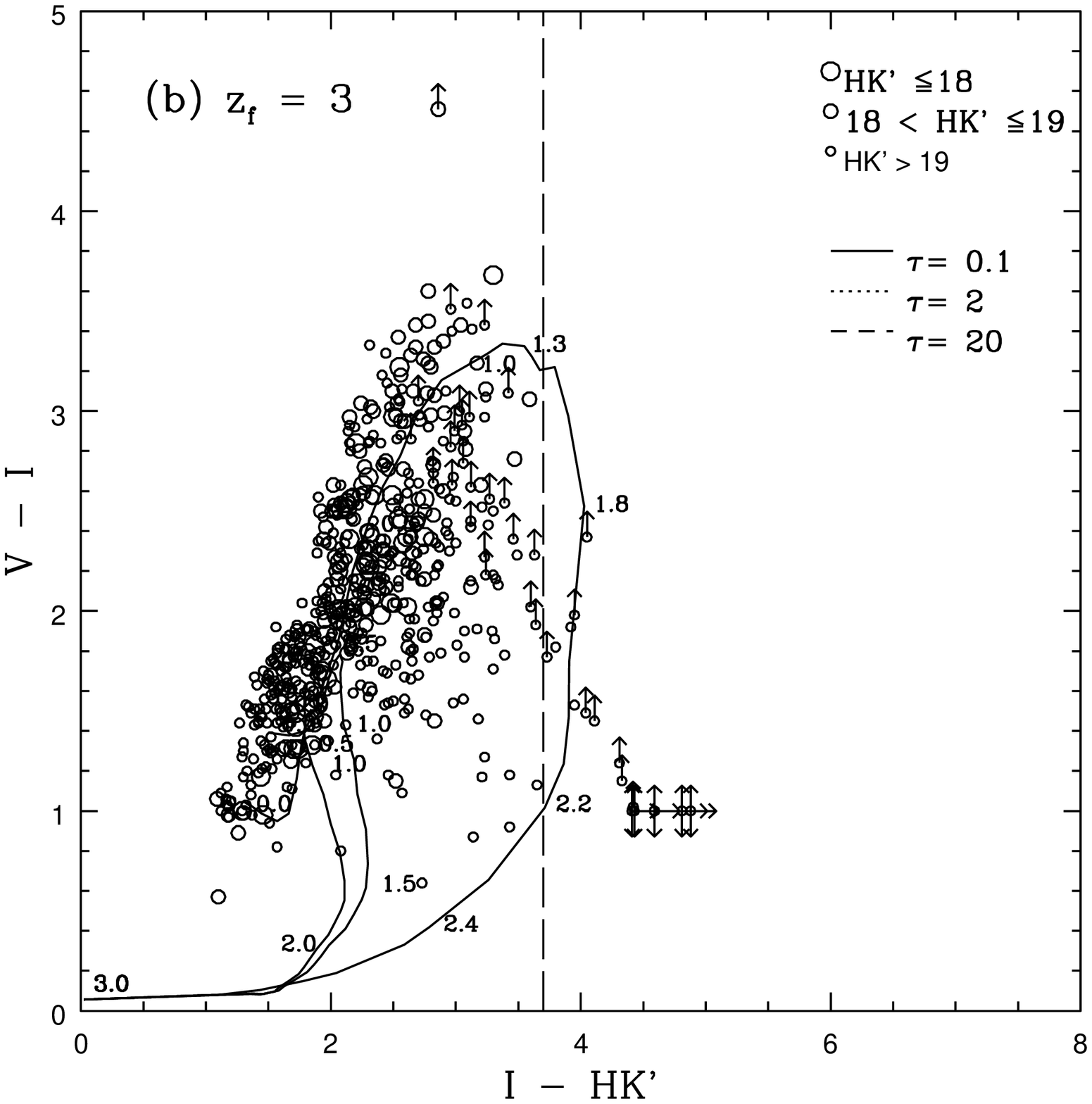,width=3.0in,height=3.0in}
\caption{$V-I$ versus $I-HK^\prime$ color-color diagrams for
the wide-field $HK^\prime\le 20.4$ sample (stars have been removed).
Bruzual \& Charlot (1996) passively evolving model tracks folded with
the appropriate $V$, $I$, and $HK^\prime$ filters are illustrated
from $0<z<3$, assuming an epoch of galaxy formation of
(a) $z_f=5$ and (b) $z_f=3$. The model tracks have
been labeled with a few representative redshifts at which the
various colors would be observed. The vertical dashed lines in the figures
illustrate the ultrared dividing line, $I-HK^\prime>3.7$.
\label{evcolcol}}
\end{figure}

In Figs.~\ref{necolcol} and \ref{evcolcol} 
we plot $V-I$ versus $I-HK^\prime$ diagrams for our galaxy samples, letting 
the size of the symbols decrease with decreasing galaxy brightness.
Galaxy evolutionary codes provide a powerful means for interpreting
galaxy color-color distributions. To aid in our interpretation of the 
data, we overlay
on the data template model galaxy tracks for three representative star 
formation histories. These models were produced with the code of
G.~Bruzual \& S.~Charlot 
(1996, private communication; hereafter BC96) assuming
solar metallicity, Salpeter initial mass function (IMF), and
exponentially declining star formation rates (SFR). The star formation
time-scales adopted are
$\tau=(0.1, 2, 20)$~Gyr, which roughly approximate elliptical,
Sa, and constant star-forming galaxies, respectively. We label the
tracks with representative redshifts over the range $0<z<3$
at which the galaxies would be observed. 
In all our model comparisons with the data
we adopt a $q_o=0.5$ cosmology and a Hubble constant of
$H_o=50\ $km s$^{-1}$ Mpc$^{-1}$.
Barring modeling uncertainties, 
galaxies following a particular evolutionary track in the color-color
plane should appear scattered about the track by
observational uncertainties. 


The model curves in Fig.~\ref{necolcol}(a,b) illustrate the no-evolution 
scenario in which spectral energy distributions (SEDs)
corresponding to local evolved galaxies 
(modeled with BC96 assuming a common epoch of formation, $z_f=5$)
are moved back in redshift. Thus,
the tracks represent color changes due to pure K-correction. In the figure
these pure K-correction models are overlaid onto the (a) wide-field and 
(b) deep-field $HK^\prime$-selected samples. Nearly all of our
galaxies in both surveys have colors consistent with a $z<1$ population. 
The deep sample also contains a substantial population of objects
consistent with high-$z$ star-forming galaxies.
Significantly, however, although there are a number of bright
objects with red $V-I$, $I-HK^\prime$ colors consistent
with the $z<1$ elliptical galaxy evolution tracks, there are no
galaxies following these tracks for $z>1$.

The only possible candidates for $z>1$ pure K-correction 
ellipticals are galaxies
found in $K$-selected field samples to have extremely red $I-K$ colors.
The $I-K$ color of a pure K-correction 
elliptical becomes very red beyond $z=1$ where
the 4000\ \AA\ break moves between the $I$ and $K$ bands.
Thereafter, the $I-K$ color is a sensitive function of the redshift.
The color of a pure K-correction elliptical galaxy at
$z=1$ is $I-K\simeq 4$. For consistency with previous surveys, we also 
select this number as the lower limit for defining 
the {\it ultrared} galaxy population. For our $HK^\prime$ sample this 
cut-off is equivalent to $I-HK^\prime=3.7$.
By $z\sim 2$ the $I-K$ color
should exceed six (i.e.\ assuming a large redshift of formation),
but observed objects this red are extremely rare. The most robust
$I-K>6$ detections are those of \markcite{hu94}Hu \&
Ridgway (1994). They found several $K$-band-bright galaxies in the field of
the quasar PC1643+4631A (most probably serendipitously discovered
field objects) with $I-K\sim 6.5$. 
Recently, \markcite{cimatti98}Cimatti et al.\ (1998)
detected one of these objects (HR10) with the submillimeter
bolometer array SCUBA on the James Clerk Maxwell Telescope, thereby
demonstrating that HR10 is a very dusty galaxy undergoing a major
episode of star formation.

\markcite{moustakas97}Moustakas et al.\ (1997) separated out a population
of $I-K\gtrsim 4$ galaxies in their sample having blue optical colors
$V-I\lesssim 2.5$ (the `red outliers') and suggested that these
objects could be fairly common. To $K\sim20$ in their sample, two secure
objects fall into this category.
For their survey area of $\sim 2$\ arcmin$^2$, this
corresponds to a surface density of $3600$\ deg$^{-2}$. For comparison, in
our wide-field sample to an equivalent $K=20.1$ ($HK^\prime=20.4$),
we find 16 objects that could
potentially satisfy the red outlier color criteria, which corresponds to
a surface density of $900^{1200}_{700}$\ deg$^{-2}$.
Given the fact that we cover a much larger area (61.8\ arcmin$^2$), it
seems likely that the Moustakas et al.\ results may have
been an upward statistical fluctuation or a clustering effect, and we
conclude that this class of galaxy is rare.

Although the ultrared galaxy population does not appear to 
correspond well with the pure K-correction elliptical track, 
this region is spanned
by some non-evolving spiral curves (e.g.\ $\tau=2$, which roughly approximates
an Sa galaxy) with $z>1$. Songaila et al.\ (1994) found no evidence
in their $K<20$ spectroscopic sample of the Hawaii Survey Fields
for any ultrared galaxies between $z=0$ and $z=1$.
They noted that their spectroscopically unidentified
objects, many of which appeared to be red $I-K$ objects, had colors
which were consistent with being primarily in the redshift interval
$z=1-2$ and falling somewhere between the color of an unevolved
Sb and an unevolved Im galaxy.


In Fig.~\ref{evcolcol}(a,b) we overlay on the wide-field data 
model curves which trace with redshift
the colors of a passively evolving population of galaxies for 
two different epochs of formation, $z_f=5$ and $z_f=3$, respectively.
Changing the cosmology to $q_o=0.15$, $H_o=65\ $km s$^{-1}$ Mpc$^{-1}$
makes very little difference in the model tracks.
The first thing we notice in this figure
is a population of objects at $I-HK^\prime>3.7$ which appears
to fall outside the range of the passively evolving model curves. 
Although this region does not seem to be well represented by the passively
evolving model curves, it may be better described by a reddened 
elliptical SED.

\begin{figure}
\centering
\hspace*{0in}\psfig{figure=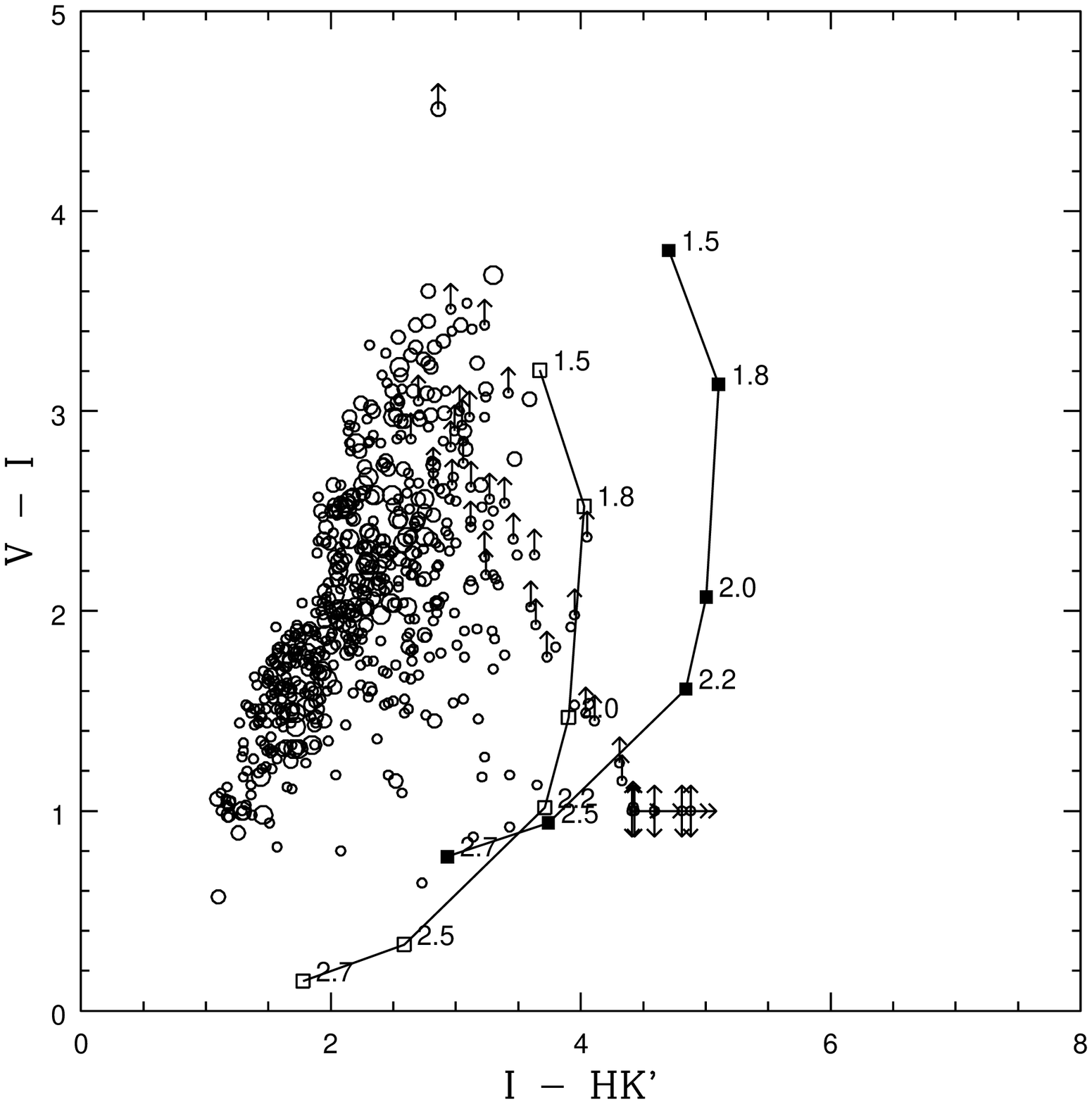,width=3.0in,height=3.0in}
\caption{The effects of reddening on the $\tau=0.1$ passive evolution 
model of BC96 in the $V-I$ versus $I-HK^\prime$ plane for $z_f=3$
using the Calzetti (1997) dust attenuation law and a color
excess of $E(B-V)=0.3$. Solid squares represent the reddened SED
colors and open squares the intrinsic SED colors. The symbols are labelled
with the redshifts of the galaxies which have those colors.
The data are from the wide-field sample.
\label{dust}}
\end{figure}

In Fig.~\ref{dust} we compare a $z_f=3$ unreddened, passively evolving
$\tau=0.1$ model galaxy with one reddened
using the dust attenuation law of \markcite{calzetti97}Calzetti (1997) for
star-forming galaxies. In the galaxy rest-frame, the reddened, 
$F_{red}(\lambda)$, and intrinsic, $F(\lambda)$, fluxes are related by
\begin{equation}
F_{red}(\lambda)=F(\lambda) 10^{-0.4 E(B-V) k(\lambda)}
\end{equation}
In Calzetti's recipe for reddening, the
attenuation of the stellar continuum, $k(\lambda)$, is given by
\begin{equation}
k(\lambda)=[(1.86-0.48/\lambda)/\lambda - 0.10]/\lambda + 1.78
\end{equation}
for $0.63\micron$ $\le\lambda\le 1.0\micron$, and
\begin{equation}
k(\lambda)=2.656(-2.156+1.509/\lambda -0.198/\lambda^2+0.011/\lambda^3) + 4.88
\end{equation}
for $0.12\micron$ $\le\lambda< 0.63\micron$. Our illustration in 
Fig.~\ref{dust} is based on a color excess of $E(B-V)=0.3$. 
The reddening law above
incorporates the effects of extinction and scattering, as well as
the geometrical distribution of the dust relative to the emitters.
We can only apply this relation over the redshift range $1.5<z<3.0$
because of her wavelength restrictions.
 

The BC96 models show that a $\tau=0.1$ star formation model will
approximate an elliptical galaxy SED after about 2~Gyr; if a galaxy 
formed at $z_f=3$ in our cosmology, it would not look like an elliptical until
a redshift of $z\sim 1.3$. It is evident from Fig.~\ref{dust} that the 
reddened elliptical model with $z_f=3$ could adequately explain the 
population of galaxies at $I-HK^\prime>3.7$. Thus, the ultrared population 
could be a population of dust-enshrouded ellipticals at $z\sim 2$ in the 
process of formation.

\section{Constraints on High Redshift Field Ellipticals}
\label{disc}

\markcite{kauffmann96}Kauffmann (1996) showed that for a standard
CDM model with density parameter $\Omega=1$ and normalization
parameter $\sigma_8=0.7$ ($b\equiv 1/\sigma_8$), the global number
density of bright elliptical galaxies at $z=1$ is smaller by a factor 
of $2-3$ than that at $z=0$. Later
\markcite{kcw96}Kauffmann, Charlot \& White (1996) applied
the Schmidt $V/V_{max}$ test to the color-selected $0.1<z<1$ early-type
population in the Canada-France Redshift Survey and found that the
observational data were {\it also} inconsistent with the standard model of
passive evolution. The density evolution observed was so strong that 
it implied that at $z=1$ only a third of the
bright early-types were already assembled and had the colors of
old, passively evolving systems. These authors
hence inferred that an increasing fraction of the
early-type population drops out of the sample with increasing
redshift, either because they are no longer single
units or because star formation has altered their colors.
\markcite{lilly95}Lilly et al.\ (1995) had
analyzed the same dataset previously, however, and had come to the
conclusion that there had been no significant evolution.

As discussed in the introduction, an alternative approach
is to place an upper limit on the number density of high-redshift 
evolved elliptical galaxies using
very deep imaging at optical and near-infrared wavelengths. 
The present very large area survey allows us to place
limits on the population of $z>1$ galaxies
having ultrared optical-near-IR colors. 
In order to compare our results with previous $K$-band surveys, 
we convert our $HK^\prime$ magnitudes to equivalent 
$K$ band magnitudes calculated from the
calibration equation $K=HK^\prime-0.30$ determined in \S\ref{obs}.
The reddest color convincingly detected in our wide-field sample is
$I-K=4.7\pm 0.3$. This is comparable to the reddest 
color reported in the $K\le 20$ sample of Cowie et al.\ (1994), 
$I-K=5.1\pm 0.4$. 
Within the 61.8\ arcmin$^2$ wide-field area we have surveyed 
($K\le 20.1$),
we detect only 16 galaxies with galaxy colors $I-K>4$.
In the 7.8\ arcmin$^2$ area of our deep HDF survey ($K\le 21.2$),
we detect 4 galaxies with colors $I-K>4$; there are, however, an 
additional 5 galaxies with only $2\sigma$ limits on their $I$ magnitudes 
which could move into the ultrared region.
In Table~\ref{tabfraction} we compare the fraction of objects (12/290)
with $I-K>4$ in the inclusive magnitude range $K=19-20$
in our wide-field 
sample and the fraction of objects (4-9/81) with $I-K>4$ in 
the inclusive magnitude range $K=20-21$
in our deep sample with the results of Cowie et al.\ (1994). 
Due to improved photometric errors and our much larger area coverage, 
our wide-field determination of the fraction of
ultrared galaxies in the range $K=19-20$ is expected to be
more reliable than that obtained from smaller fields by
Cowie et al.\ (1994). Our improved $I$-band data may also be 
responsible for the fact that we find a smaller fraction of ultrareds 
in our small field $K=20-21$ sample than Cowie et al., although
this discrepancy could also be a galaxy clustering effect.


\begin{deluxetable}{c c c c c}
\tablecolumns{5}
\tablewidth{0pc}
\tablenum{1}
\tablecaption{Fraction of $I-K>4$ Galaxies}
\tablehead{
\colhead{} & \multicolumn{2}{c}{Current Paper}
&\multicolumn{2}{c}{Cowie et al.\ (1994)} \\
\colhead{$K$ (mag)} & \colhead{Fraction} & \colhead{$\pm 1\sigma$ Range} &
\colhead{Fraction} & \colhead{$\pm 1\sigma$ Range}
}
\startdata
19-20 & 0.041 & 0.030-0.057 & 0.07 & 0.025-0.14 \nl
20-21 & 0.049-0.11 & 0.026-0.16 & 0.20 & 0.15-0.27 \nl
\enddata
\label{tabfraction}
\end{deluxetable}

We can use our data to
place a limit on the number of present-day elliptical galaxies 
that could have formed in an early burst at high redshift. To do this we
need to calculate the space density of ultrared galaxies
in our wide-field sample and compare it with the space density of 
local ellipticals to the same magnitude limit. We
assume the ultrared galaxies in our wide-field survey are
distributed within the redshift range $1.3<z<2.2$, corresponding to a
late formation $z_f=3$ (Fig.~\ref{evcolcol}b), in order to estimate the
volume for our assumed cosmology ($q_o=0.5$ with 
$H_o=50\ $km s$^{-1}$ Mpc$^{-1}$). A higher $z_f$ would give a 
larger volume and hence a tighter constraint.
We find the space density, $n_{ultrared}=N_{obs}/V$, 
of the ultrared population to be 
$n_{ultrared}\simeq 1\times 10^{-4} h_{50}^3\,$Mpc$^{-3}$
(where $h_{50}$ is a dimensionless quantity, $h_{50}=H_o/50$).
This value is to be compared with the local number density, 
given by the relation
\begin{equation}
n=\phi^*_{50} {h^3_{50}} \Gamma(1+\alpha,\beta)
\label{neq}
\end{equation}
where $\Gamma(1+\alpha,\beta)$ is the incomplete gamma function with
\begin{equation}
\beta=10^{-0.4(M_{min} - M^*)}
\label{betaeq}
\end{equation}
(\markcite{yoshii88}Yoshii \& Takahara 1988) and $\alpha$ the faint-end
slope parameter of the Schechter luminosity function 
\markcite{schechter76}(Schechter 1976).
At the limit of our wide-field sample, $K=20.1$, a galaxy
at a mean redshift of $\bar{z}=1.75$ would have an absolute magnitude
$M_{min}=M_K=-23.4 + 5\log h_{50}$, after including passive
evolution and K-corrections derived from a BC96 $\tau=0.1$ model
($z_f=3$, $q_o=0.5$). At present no local $K$-band luminosity functions 
by morphological type have been determined for the field. Therefore, 
we convert the $B$-band pure-elliptical luminosity function
parameterization of
\markcite{marzke94}Marzke et al.\ (1994) ($M_B^*=-19.23 + 5\log h_{100}$,
$\alpha=-0.85$, and $\phi^*_{100}=(1.5\pm 0.4)\times 10^{-3}\,$Mpc$^{-3}$)
into a $K$-band elliptical 
luminosity function using the $B-K$ color of an elliptical galaxy
at $z=0$ ($B-K=4.43$; \markcite{huang98}Huang et al.\ 1998). This gives $M_K^*=-25.16+5\log h_{50}$
and $\phi^*_{50}=1.9\times 10^{-4}\,$Mpc$^{-3}$. 
Using Eqs.~\ref{neq} and \ref{betaeq}, we
estimate the local elliptical galaxy space density to be
$n_{local}\simeq 2\times 10^{-4} h_{50}^3\,$Mpc$^{-3}$. 
These results indicate that the space density of the 
ultrared population is $\sim 50$ per cent of the space density 
of the local field elliptical population with $M_K<-23.4$.

It is important to be aware of possible sources of systematic errors
in this calculation. First,
the comoving volume needed for the ultrared number density
calculation depends on the model estimates of the redshift range over
which $I-K>4$; the volume would be larger and $n_{ultrared}$
correspondingly smaller for $z_f>3$. Second,
the normalization of the density of the present-day luminosity function
is not fully settled, with other surveys giving values that differ
from the one we have used by $\pm 25$ per cent
(see, e.g., the discussion in \markcite{huang98}Huang et al.\ 1998).
Third, a comparison with a local $K$-band pure-elliptical luminosity 
function would be preferable to the above conversion of a $B$-band 
pure-elliptical luminosity function; such a luminosity function 
should soon be available (J.-S.~Huang, private communication). 
Fourth, the fraction of ultrareds 
is mildly dependent on cosmology; the above calculation, when 
repeated for a $q_o=0.15$ cosmology with $H_o=65\ $km s$^{-1}$ Mpc$^{-1}$, 
gives about a 10\ per cent lower ratio of $n_{ultrared}/n_{local}$, so our
$q_o=0.5$ calculation is conservative in this regard.
Fifth, the possibility of reddening in the high redshift sample should 
also be considered. If we assume that the possible reddening discussed in 
\S\ref{ultrared}
applies to the entire ultrared population but not to the local population, 
then we find that the ratio increases by about 15\ per cent. Finally, 
although our results are from a large area survey, they are still based on 
a single field. This introduces a statistical uncertainty associated 
with field-to-field variations, particularly 
in a population as small as the ultrared population. Results of other 
upcoming deep $K$-band surveys (e.g.\ papers in preparation by 
McCracken et al., Eisenhardt et al., and Barger et al.) will be useful in
examining this issue.

\section{Summary}
\label{summary}

We have presented wide-field $V$, $I$, and $HK^\prime$ images of the Hubble
Deep Field and Flanking Fields, as well as a deeper $HK^\prime$ image of
the Hubble Deep Field. Most of the objects in our sample fall within the
expected regions of the $V-I$ versus $I-HK^\prime$ color-color plane for blue
galaxies at a range of redshifts or red elliptical galaxies at $z<1$.
Significantly, no galaxies are found to follow the track of a pure
K-correction elliptical at $z>1$ in either of our two samples. 
The track of a passive evolution model with a redshift of formation
$z_f=3$ roughly describes the envelope of the galaxy population in the
$V-I$ versus $I-HK^\prime$ plane, except for a very small population of red 
outliers that could be dusty ellipticals in the process of formation.
To an equivalent $K=20.1$ limit over a large 61.8\ arcmin$^2$ area, we 
find a surface density for $I-K>4$ objects of
$900^{1200}_{700}$\ deg$^{-2}$, which indicates that this type of object
is even more rare than previous small area surveys have suggested.
Since an $I-K>4$ color could be mimicked by a highly-reddened low-redshift 
galaxy, this is effectively an upper limit on the population of 
$z>1$ ultrared galaxies, provided that there is not a population of
completely dust-obscured high redshift galaxies. We find that the
space density of the ultrared galaxies is only a fraction of
the estimated space density of present-day field ellipticals with
$M_K<-23.4$, and hence 
we infer that not all local field ellipticals could have formed in 
single bursts at high redshift.

\newpage

\begin{acknowledgements}

We thank William Vacca for helpful interactions during the course of this 
work and Richard Ellis, Roberto Abraham, and Felipe Menanteau for 
discussions on related research. We thank the anonymous referee for
helpful comments.
This work was supported by grant AR06377.06-94A from Space Telescope 
Science Institute, which is operated by AURA, Inc., under NASA 
contract NAS 5-26555. AJB also acknowledges support from NASA through 
contract number P423274 from the University of Arizona, under NASA grant 
NAG5-3042. 

\end{acknowledgements}


\begin{references}
 
\reference{alfonso93}
{Arag{\'o}n-Salamanca, A., Ellis, R.S., Couch, W.J., Carter, D. 1993, \mnras, 
262, 764}
 
\reference{baugh96}
{Baugh, C.M., Cole, S., Frenk, C.S. 1996, \mnras, 283, 1361}
 
\reference{bertin96}
{Bertin, E., Arnouts, S. 1996, A\&AS, 117, 393}

\reference{calzetti97}
{Calzetti, D. 1997, in The Ultraviolet Universe at Low and High Redshift,
eds. Waller, H., Fanelli, M.N., Hollis, J.E., Danks, A.C., AIP. Conf.
Proceedings 408, 403}

\reference{cimatti98}
{Cimatti, A., Andreani, P., R\"ottgering, H., \& Tilanus, R. 1998,
Nature, 392, 895}

\reference{cohen96}
{Cohen, J.G. et al.\ 1996, \apjl, 471, L5}
 
\reference{cowie94}
{Cowie, L.L., Gardner, J.P., Hu, E.M., Songaila, A., Hodapp, K.-W.,
Wainscoat, R.J. 1994, \apj, 434, 114}
 
\reference{eggen62}
{Eggen, O.J., Lynden-Bell, D., Sandage, A.R. 1962, \apj, 136, 748}
 
\reference{ellis97}
{Ellis, R.S. et al.\ 1997, \apj, 483, 582}
 
\reference{fomalont97}
{Fomalont, E.B., Kellermann, K.I., Richards, E.A., Windhorst, R.A., 
Partridge, R.B. 1997, \apjl, 475, 5}

\reference{frances98}
{Franceschini, A. et al.\ 1998, astro-ph/9806077}

\reference{gardner92}
{Gardner, J.P. 1992, Ph.D. thesis, Univ. Hawaii}

\reference{hodapp96}
{Hodapp, K.-W. et al.\ 1996, NewA, 1, 177}
 
\reference{hogg97}
{Hogg, D.W., Neugebauer, G., Armus, L., Matthews, K., Pahre, M.A. 1997,
\aj, 113, 2338}
 
\reference{hu94}
{Hu, E.M., Ridgway, S.E. 1994, \aj, 107, 1303}

\reference{huang97}
{Huang, J.-S., Cowie, L.L., Gardner, J.P., Hu, E.M., Songaila, A.,
Wainscoat, R.J. 1997, \apj, 476, 12}

\reference{huang98}
{Huang, J.-S., Cowie, L.L., Luppino, G.A. 1998, \apj, 496, 31}
 
\reference{kaiser97}
{Kaiser, N. 1997,
http://tiro.ifa.hawaii.edu/~kaiser/imcatdoc/mainindex.html}

\reference{kauffmann96}
{Kauffmann, G. 1996, \mnras, 281, 487}
 
\reference{kauffcharlot98}
{Kauffmann, G., Charlot, S. 1998, \mnras, 294, 705}
 
\reference{kcw96}
{Kauffmann, G., Charlot, S., White, S.D.M. 1996, \mnras, 283, 117}
 
\reference{kauffmann93}
{Kauffmann, G., White, S.D.M., Guiderdoni, B. 1993, \mnras, 264, 201}
 
\reference{kelson97}
{Kelson, D.D., van Dokkum, P.G., Franx, M., Illingworth, G., Fabricant, D.
1997, \apj, 478, L13}

\reference{kron80}
{Kron, R.G. 1980, \apjs, 43, 305}
 
\reference{lilly95}
{Lilly, S.J., Tresse, L., Hammer, F., Crampton, D., Le F{\`e}vre, O. 1995,
\apj, 455, 108}
 
\reference{lowenthal97}
{Lowenthal, J. et al.\ 1997, \apj, 481, 673}

\reference{marzke94}
{Marzke, R.O., Geller, M.J., Huchra, J.P., Corwin, Jr. H.G. 1994, \aj, 108, 437}
 
\reference{moustakas97}
{Moustakas, L.A., Davis, M., Graham, J.R., Silk, J., Peterson, B.A.,
Yoshii, Y. 1997, \apj, 475, 445}
 
\reference{moustakas98}
{Moustakas, L.A., Davis, M., Zepf, S.E., Bunker, A.J., in
The Young Universe: Galaxy Formation and Evolution at Intermediate and
High Redshift, eds. S. D'Odorico, A. Fontana, E. Giallongo, A.S.P. Conf.
Ser., 1998, astro-ph/9712135}
 
\reference{phillips97}
{Phillips, A.C. et al.\ 1997, \apj, 489, 543}

\reference{schechter76} 
{Schechter, P.L. 1976, \apj, 203, 297}
 
\reference{songaila97}
{Songaila, A. 1997, http://www.ifa.hawaii.edu/~cowie/tts/tts.html}

\reference{songaila94}
{Songaila, A., Cowie, L.L., Hu, E.M., Gardner, J.P. 1994, \apjs, 94, 461}

\reference{stan95}
{Stanford, S.A., Eisenhardt, P.R.M., Dickinson, M. 1995, \apj, 450, 512}
 
\reference{stan98}
{Stanford, S.A., Eisenhardt, P.R., Dickinson, M. 1998, \apj, 492, 461}
 
\reference{steidel96}
{Steidel, C.C., Giavalisco, M., Dickinson, M., Adelberger, K. 1996,
\aj, 112, 352}
 
\reference{tinsley76}
{Tinsley, B.M., Gunn, J.E. 1976, \apj, 203, 52}
 
\reference{vDF97}
{van Dokkum, P.G., Franx, M. 1996, \mnras, 281, 985}
 
\reference{white91}
{White, S.D.M., Frenk, C.S. 1991, \apj, 379, 52}
 
\reference{williams96}
{Williams, R.E., et al. 1996, \aj, 112, 1335}

\reference{yoshii88}
{Yoshii, Y., Takahara, F. 1988, \apj, 326, 1}
 
\reference{zepf98}
{Zepf, S. 1997, Nature, 390, 377}
 
\reference{bzb97}
{Ziegler, B., Bender, R. 1997, \mnras, 291, 527}
 
\end{references}
\end{document}